\documentclass{aa}

\usepackage{graphicx}
\usepackage{natbib}
\usepackage{amssymb}
\usepackage{txfonts}

\bibpunct{(}{)}{;}{a}{}{,}
\begin{document}

\title{ \ion{H}{ii} regions in symbiotic binaries and their radio emission}

\author{A. J. Gawryszczak \and
  J. Miko\l ajewska \and
  M. R\'o\.zyczka 
}

\offprints{A. J. Gawryszczak, \email{gawrysz@camk.edu.pl}}

\institute{
  Nicolaus Copernicus Astronomical Center, Bartycka 18, Warsaw, PL-00-716, Poland
}

\date{Received (date); accepted (date)}

\abstract{The slow and dense wind from a symbiotic red giant (RG)
can be significantly deflected toward the orbital plane by the gravitational 
pull of the companion star. In such an environment, the ionizing radiation 
from the companion creates a highly asymmetric \ion{H}{ii} region. We present 
three-dimensional models of \ion{H}{ii} regions in symbiotic S-type systems, 
for which we calculate radio maps and radio spectra. We show that the 
standard assumption of spherically symmetric RG wind results in wrong shapes, 
sizes and spectra of ionized regions, which in turn affects the observational 
estimates of orbital separation and mass loss rate. A sample of radio maps 
and radio spectra of our models is presented and the results are discussed in 
relation to observational data. 
}

\maketitle

\section{Introduction}
\label{Intro}
Symbiotic stars are binary systems consisting of a red giant (RG) as
the primary, and a post-AGB star or MS-dwarf as the secondary. The RG
primary is the source of a high density wind accreted by the
secondary. In closer binaries ($a < 10\,\mathrm{AU}$) the wind is
significantly deflected toward the orbital plane by the gravitational
pull of the secondary
\citep[see][ and references therein]{02GMR}. This process results in an
enhanced density region at the orbital plane, with a disk-like density
maximum surrounding the secondary.

Radio emission from symbiotic stars is dominated by {\it ff} radiation
from the ionized gas. The {\it ff} spectrum of an \ion{H}{ii} region
consists of three parts: optically thick at low frequencies (with
spectral index $\alpha\ge0.6$), optically thin at high frequencies
(with $\alpha\approx-0.1$), and optically semi-thick at intermediate
frequencies around at a turnover frequency $\nu_\mathrm{t}$, with
$\alpha$ smoothly decreasing from its optically thick value to
$-0.1$. \cite{84TS} show that orbital separation $a$, ionizing
luminosity $L_\mathrm{ph}$, and the ratio of mass-loss rate to wind
velocity $\frac{\dot{M}}{v}$ may be derived from such observables as
turnover frequency, monochromatic flux at that frequency
$S_\mathrm{t}$, and the value of the optically thick spectral index.

However, their calculations are based on the spherical wind model
originally developed by
\cite{1984STB} (hereafter STB), which, as we mentioned above, is
incorrect in closer systems. Good examples of such systems are
symbiotic S-type systems, containing a normal M giant and having
orbital periods of order a few years
\citep[e.g.][ and references therein]{bm00}.

Recently, \cite{01MI} have obtained for the first time the spectral
energy distribution (SED) at wavelengths between $6\,\mathrm{cm}$ and
$0.85\,\mathrm{mm}$ for one of the prototypical S-type systems,
\object{CI Cyg}, during quiescence. Their data allowed to determine
$\nu_\mathrm{t}$, the optically thin {\it ff} emission measure, from which
a lower limit to $L_\mathrm{ph}$ and an estimate of $a$ were
calculated within the STB framework. Unfortunately, the comparison of
these estimates with the known orbital and stellar parameters of
\object{CI Cyg} poses a serious problem for the STB model. In
particular, $a$ is overestimated by a factor of $\sim 35$, whereas
$L_\mathrm{ph}$ (whose value is known from optical/UV \ion{H}{i}
emission lines and \ion{He}{ii} {\it bf+ff} continuum) is
underestimated by a factor of $\sim 20$.

The above inconsistency strongly indicates that the STB model needs
revision. The most obvious improvements are related to the asphericity
of the wind, as suggested both by numerical work
\citep[see][ and references therein]{02GMR} and observations. From the
observational point of view there is little doubt about the origin of
the ionized medium. The mm radio emission shows some correlation with
the mid-IR flux, and the radio luminosity increases with the
$\mathrm{K}-[12]$ colour. This indicates that warm dust is involved in
the mass flow along with the ionized gas, and suggest that the red
giant is the source of this material \citep{mio02}. Such a picture is
also supported by the spectral analysis of symbiotic nebulae which
showed CNO abundances similar to those observed in normal red giants
\citep{88NSVS} Optical and radio imaging of symbiotic stars reveal
aspherical nebulae, often with bipolar lobes and jet-like components
\citep{99cfsbg}. It is hard to explain how such structures might
develop if the RG wind were spherical. Additional, indirect evidence
for the asphericity of the RG wind comes from emission line
profiles. It is based on the stationary, blueshifted
$\mathrm{H}_\alpha$ absorption originating close to the orbital plane
and indicating that the ionized region is probably bounded on all
sides by a significant amount of dense neutral material \citep[see the
discussion in][]{02GMR}.

In the present paper we account for the asphericity of the RG as
resulting from the interaction with the secondary. With the help of
3-D hydrodynamical models we demonstrate that
\ion{H}{ii} regions ionized by the secondary in S-type systems must
indeed have shapes, sizes and spectra significantly different from
those obtained by STB. A representative sample of radio spectra and
radio maps of our models is also generated.

In Sect.~\ref{methods} we discuss the numerical techniques used in this work
and details of the physical setup of our models. The models are
presented in Sect.~\ref{results}. In Sect.~\ref{discussion} the results of
our simulations are
compared to observed symbiotic systems and a brief summary of our work is
provided.

\section{Methods and assumptions}
\label{methods}
The aim of the present paper is to study the properties of \ion{H}{ii}
regions in symbiotic S-type stars. In such systems the \ion{H}{ii}
region develops in a dense RG wind whose density distribution is
modified by the gravitational pull of the secondary. As a highly
aspherical distribution may result even in cases when the wind is
originally spherically symmetric, a multidimensional approach to the
problem is necessary.

We begin with 3-D hydrodynamical simulations of RG winds interacting
with the companion to the mass-losing star. The simulations are
performed with the help of the Smoothed Particle Hydrodynamics (SPH)
technique. An inertial coordinate system is used, with the origin at
the mass center of the binary.

Our implementation of the SPH is based on variable smoothing length
with spline kernel introduced by \cite{85ML}. The number of
neighbouring particles is set to~40. For practical reasons (CPU
time), the total number of particles in the computational domain is
limited to $10^5$. Whenever this limit is exceeded, those most distant
from the system are removed from the domain. The distance beyond which
removals occur ($r_\mathrm{rem}$) stabilizes within several orbital
periods of the binary, and an almost stationary rotating density
distribution is obtained. The validity of such an approach was
demonstrated by \cite{02GMR}. With the above limit imposed on the
number of particles, $r_\mathrm{rem}$ does not exceed
$60\,\mathrm{AU}$ (see Table~\ref{ModParam}).

Within $0.1\,\mathrm{AU}$ from the secondary
we remove all particles with velocities directed toward the secondary. 
This procedure is applied to avoid prohibitively short time-steps.

There are good reasons to believe that RG winds are intrinsically aspherical 
\citep[see e.g.][ and references therein]{frtyl,96DH,00RDH}. The
asphericity may result from internal processes taking place in the
star and/or presence of the secondary. However, since the theory is
not yet advanced enough for precise quantitative predictions, these
effects are not included in our models. In particular, we assume that
although the RG rotates synchronously it does not fill its Roche lobe,
so that the mass flow through the inner Lagrangian point can be
neglected. In effect, the wind is represented by a stream of particles
launched at a constant rate from points distributed randomly and
uniformly on the RG surface. The initial temperature of the wind is
$3000\,\mathrm{K}$, and the initial velocity of wind particles is set
to $1\,\mathrm{km\,s^{-1}}$ ({ note that, as the wind propagates across 
the grid its velocity increases; a measure of the terminal wind velocity 
is introduced and explained in the following paragraphs}). 
Both temperature and velocity of the wind
are constant on the surface of the star.

Further, we assume that the wind is composed of dust and ideal
monoatomic gas which are dynamically coupled (i.e. they move at the
same velocity). The details of wind acceleration process are not
followed. Instead, we assume that the gravity of the red giant is
balanced by radiation pressure on the dust, so that it does not appear
explicitly in our equations. Thus, if the RG were single, the only
force acting on the wind would be due to gas-pressure gradients,
whereas in a binary the wind is additionally subject to the gravity of
the companion. Such an approximation may seem rather crude; however
\cite{02GMR} demonstrated that it leads to terminal velocities
comparable to those obtained by \cite{00WBJHS} for winds in which
radiation pressure nearly balances the gravity of the wind source.

Finally, we neglect radiation pressure from the secondary ({  note that 
in the case of a white dwarf,for any reasonable grain size the force due to 
radiation pressure on dust grains is much smaller than gravity}).
Self-gravity of the wind and any explicit radiative heating or
cooling are also neglected. The effects of cooling are accounted for by a 
nearly isothermal equation of state employed for the wind ($p = \kappa
\rho^\gamma$, with $\gamma = 1.01$). {  A likely source of strong 
nonaxisymmetric effects is magnetic field, as the MHD collimation may be much 
more efficient than the gravitational focusing discussed here 
\citep[see][ and references therein]{99GLR,2000GL,2001GLF}. However, there is 
no information available about magnetic fields in symbiotic systems and we 
feel that it is too early to incorporate them into models: the number of free 
parameters would simply grow too large.
}

\section{Results} 
\label{results}

All binaries considered here consist of a $1.0\,M_{\sun}$ red giant and a 
$0.6\,M_{\sun}$ white dwarf. The orbital separation $a$ varies between 1 and 
$4\,\mathrm{AU}$, which is representative of S-type symbiotic systems. 
Following \cite{mio02} a mass loss rate of 
$\dot{M}= 10^{-7}\,M_{\sun}\,\mathrm{yr}^{-1}$ is adopted for the RG wind, 
while the ionizing luminosity of the secondary $L_\mathrm{ph}$ varies 
between $4\cdot10^{44}$ and $8\cdot10^{45}\,\mathrm{phot\,s^{-1}}$. The 
latter range is compatible with estimates based on mm/submm observations 
\citep{95is,01MI}, but is a factor of $\sim$10 smaller than estimates based 
on studies of optical/UV \ion{H}{i} and \ion{He}{ii} emission lines and 
{\it bf+ff} continuum \citep{01MI,mio02}.

The choice of the lower estimate is dictated by the size of our
hydrodynamical wind models ($60\div100\,\mathrm{AU}$ in diameter),
which at $L_\mathrm{ph}\ga10^{46}$ would be practically entirely
ionized. The parameters of symbiotic binary models obtained in this
work are listed in Table~\ref{ModParam} ($v_\mathrm{av}$ in column~3
is a measure of the wind velocity; see Section~\ref{shapes} for
an explanation). The radius of the red giant is in all cases equal to
$0.5\,\mathrm{AU}$. A simple method of scaling our results to
combinations of $\dot{M}$ and $L_\mathrm{ph}$ not covered by actual
calculations is given in Section~\ref{spectra}.

\subsection{RG wind models}
\label{models}
\begin{figure*}
  \centering
  \raisebox{.39\hsize}{\parbox{1em}{A}}
  \resizebox{.85\hsize}{!}{
    \includegraphics{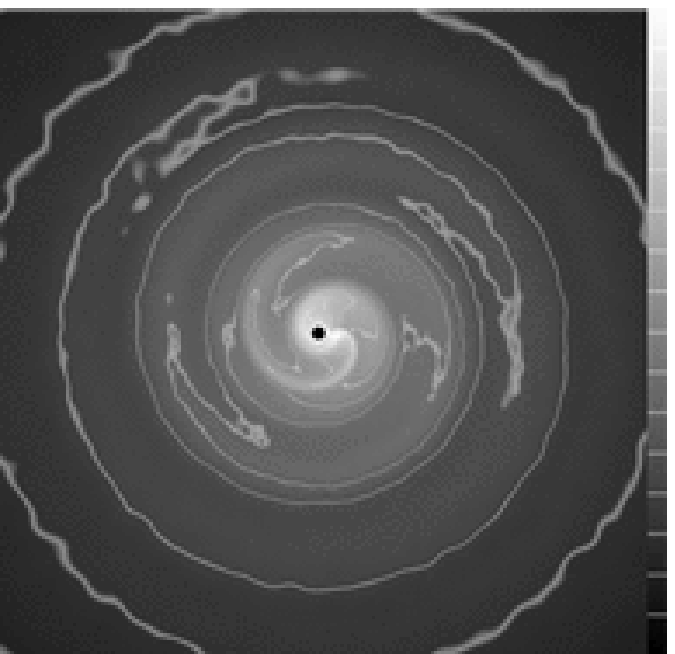}
    \includegraphics{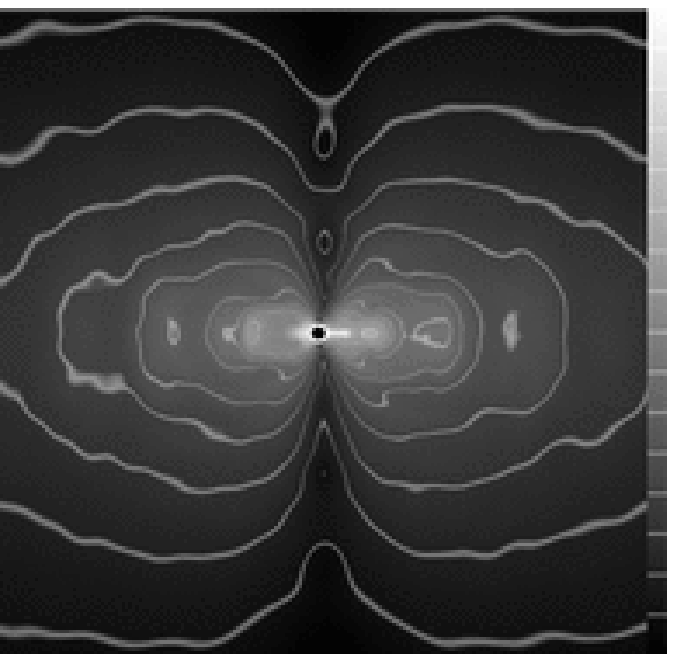}
  }\\
  \raisebox{.39\hsize}{\parbox{1em}{B}}
  \resizebox{.85\hsize}{!}{
    \includegraphics{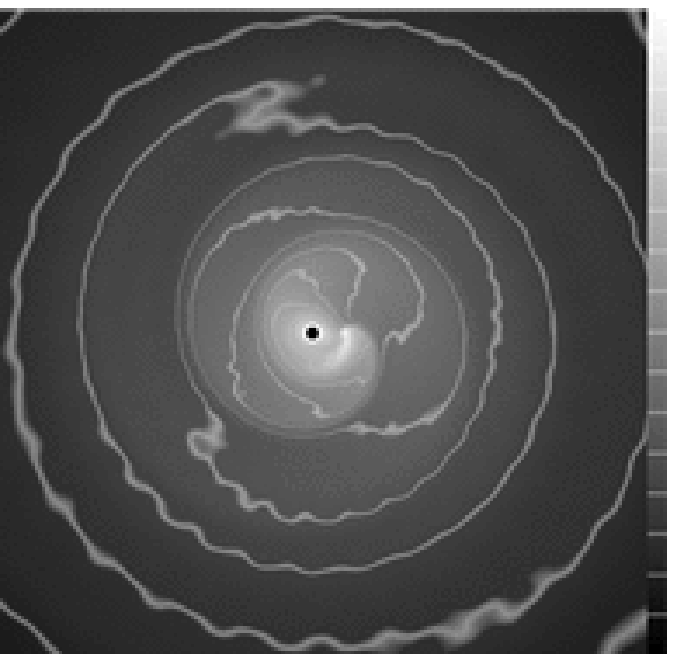}
    \includegraphics{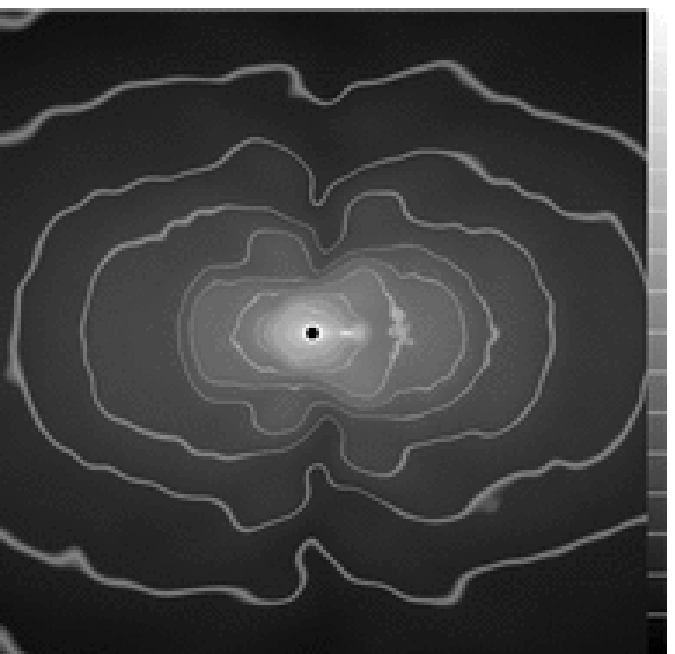}
  }\\
  \raisebox{.39\hsize}{\parbox{1em}{C}}
  \resizebox{.85\hsize}{!}{
    \includegraphics{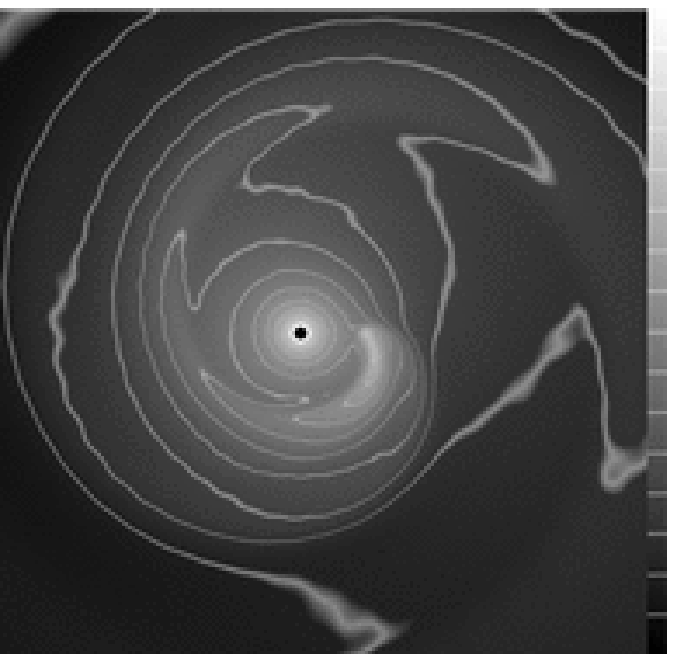}
    \includegraphics{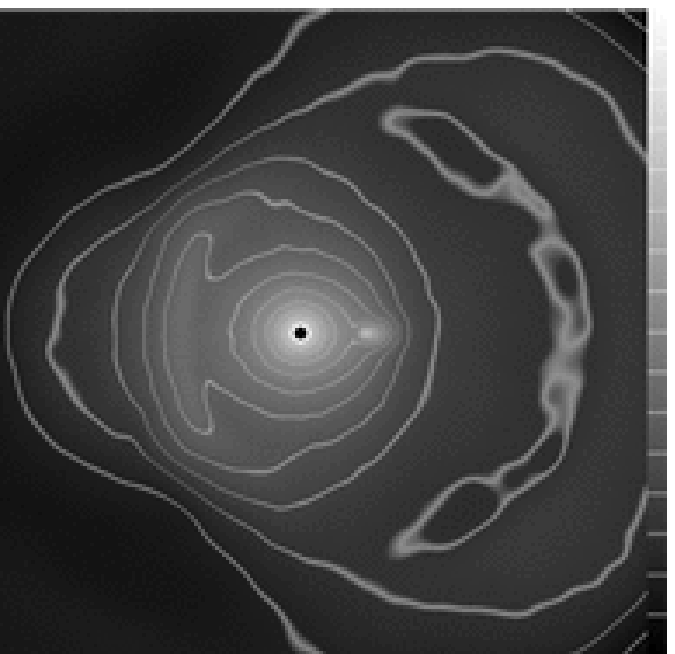}
  }\\
  \hspace{1em}
  \parbox{.425\hsize}{\center \small orbital plane }
  \parbox{.425\hsize}{\center \small perpendicular to orbital plane }
  \caption{Density distributions of models A, B and C. The logarithmic density 
    scale spans 5 orders of magnitude. In each frame a region of
    $40\times40\,\mathrm{AU}$ is shown. The system rotates counterclockwise.}
  \label{fig:dens_distr}
\end{figure*}

For each binary in Table~\ref{ModParam} we integrate the equations of
hydrodynamics until the density distribution of the RG wind relaxes to
a stationary state. Density distributions of the relaxed models are
shown in (Fig.~\ref{fig:dens_distr}). The general effect of the
secondary is to deflect the RG wind toward the orbital plane, and as a
result the relaxed density distributions are flattened. This flattening
weakens with increasing orbital separation because of the increasing ratio
of the wind velocity at the secondary's orbit to the orbital velocity of the
secondary. However, all models develop a high density region in the
form of a disk around the secondary.

\begin{table}[h]
  \caption[]{List of models}
  \label{ModParam}
  $$ 
  \begin{array}{lllllll}
    \hline
    \noalign{\smallskip}
    \mathrm{Model} & a\ \mathrm{[AU]} & v_\mathrm{av}\ \mathrm{[km\,s^{-1}]} & r_\mathrm{rem}\ \mathrm{[AU]} & \mathrm{L_\mathrm{ph}}\ \mathrm{[phot\,s^{-1}]}\\
    \noalign{\smallskip}
    \hline
    \noalign{\smallskip}
    \mathrm{A} & 1.0 & 25.0 & 33 & 7.3\cdot10^{44}\div 7.3\cdot10^{45}\\ 
    \mathrm{B} & 2.0 & 17.0 & 43 & 8.0\cdot10^{44}\div 8.0\cdot10^{45}\\
    \mathrm{C} & 4.0 & 17.0 & 54 & 4.0\cdot10^{44}\div 4.0\cdot10^{45}\\
    \noalign{\smallskip}
    \hline
  \end{array}
  $$ 
\end{table}

All models also display a three-dimensional spiral structure resulting
from the shock wave excited by the motion of the secondary across the
RG wind. The spirals are a very general feature of slow winds in
binary systems, and they were also observed in simulations reported by
\cite{98MM1,99MM2} or \cite{02GMR}. As one might expect, in closer
binaries the spiral is more tightly wound than in the wider ones.

In the closest binary funnel-like density minima oriented
perpendicularly to the orbit are also observed. Two effects contribute
to the formation of these funnels. First, in a close binary the wind
is emitted from a source moving with a high orbital velocity, and as
such it carries appreciable angular momentum. Away from the orbital
plane the centrifugal force associated with that momentum causes the
wind gas to clear the vicinity of the polar axis. Secondly, and
unlike that of the primary, the gravity of the secondary is not
balanced by radiation pressure, and the wind flowing perpendicularly
to the orbit is retarded.

Below we argue that all these features (flattening, spirals and
funnels) are important factors influencing the observational
properties of the \ion{H}{ii} regions created in the RG wind by the
ionizing radiation from the secondary.

\subsection{Shapes and sizes of \ion{H}{ii} regions} 
\label{shapes}

\begin{figure}
  \centering
  \resizebox{\hsize}{!}{\includegraphics{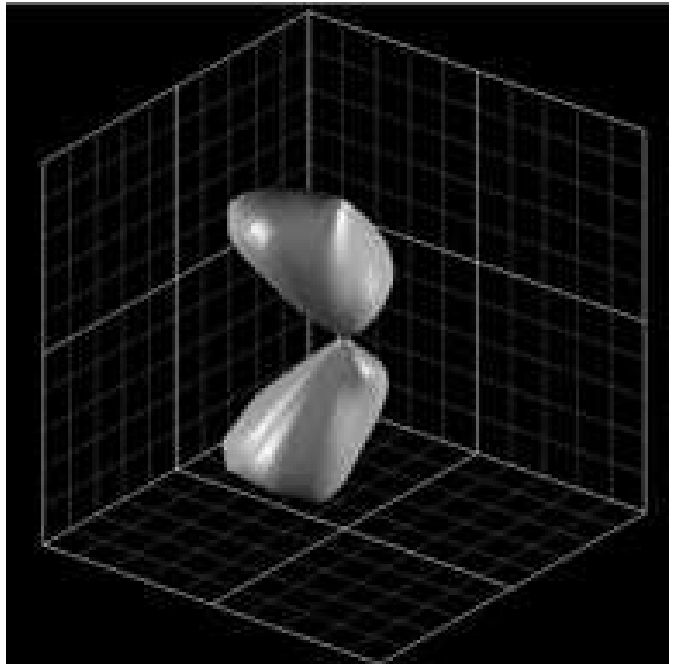}}
  \caption{3D view of the \ion{H}{ii} region boundary in model B with
    $L_\mathrm{ph}=2.4\cdot10^{45}$. A region of 
    $10\times10\times10\,\mathrm{AU}$ is shown.}
  \label{fig:hii_3d}
\end{figure}

Once the RG wind has been relaxed, its density distribution is
smoothed by averaging over several tens of consecutive time steps. The
aim of this procedure is to reduce the numerical noise resulting from
the "grainy" nature of the SPH models. The smoothed density is mapped
onto a spherical grid centered on the secondary, with grid points
spaced uniformly in angular coordinates and concentrated at the
secondary in the radial coordinate. Then, the source of ionizing
photons is turned on at the location of the secondary, and the
boundary of the \ion{H}{ii} region is found in a local Str\"omgren
sphere approximation (i.e. Str\"omgren radii are calculated for all
directions defined by grid points in angular coordinates). An example
result of this procedure is shown in Fig.~\ref{fig:hii_3d}. Separate
treatment of ionization and dynamics is justified by the fact that the
recombination time is one order of magnitude shorter than the orbital
period at the edge of our computational domain, and more than 3 orders of
magnitude near the secondary.

In the approach developed by STB, the shape and size of the \ion{H}{ii}
region are uniquely determined by the value of the parameter $X$,
which is a combination of the mass loss rate, wind velocity, orbital
separation, and ionizing luminosity:
\begin{equation}
  X=\frac{4\pi \mu^2 m_\mathrm{H}^2}{\alpha_\mathrm{rec}} a L_\mathrm{ph} 
    \left(\frac{v_\mathrm{w}}{\dot{M}}\right)^2,
\end{equation}
where $\alpha_\mathrm{rec}$ and $v_\mathrm{w}$ are recombination
coefficient and wind velocity (which is constant in space and time),
while $\mu$ and $m_\mathrm{H}$ have their standard meaning.

While STB assume a wind of constant velocity, in our models the flow
accelerates with the distance from the RG, and the acceleration
proceeds at a variable (location-dependent) rate. This is due partly
to thermal pressure gradients, and partly to the centrifugal force
associated with the angular momentum transferred to the wind from the
secondary; see
\cite{02GMR}. In order to calculate $X$ from our models, we use a
density-weighted velocity $v_\mathrm{av}$, averaged over the whole
solid angle (the averaging is performed near the edge of the
computational domain). For each binary a range of \ion{H}{ii} regions
with different $X$--values may then be easily generated by allowing
for variations in $L_\mathrm{ph}$ (alternatively, $\dot{M}$ may be
scaled by multiplying the density of the RG wind in each grid cell by
the same factor).

To obtain the STB counterpart of our model, for the same $\dot{M}$ a
density distribution of the spherical wind from the primary is
calculated, assuming $v_\mathrm{w}=v_\mathrm{av}$. An appropriate part
of that distribution is fed into the spherical grid centered on the
secondary, and the same Str\"omgren-routine is employed to find the
boundary of \ion{H}{ii} region. We checked that the boundary found
this way agrees to within a few percent with the boundary derived from
equation (1) of STB.

\begin{figure*}
  \centering
  \resizebox{\hsize}{!}{
    \raisebox{.23\hsize}{\parbox{1em}{A}}
    \includegraphics{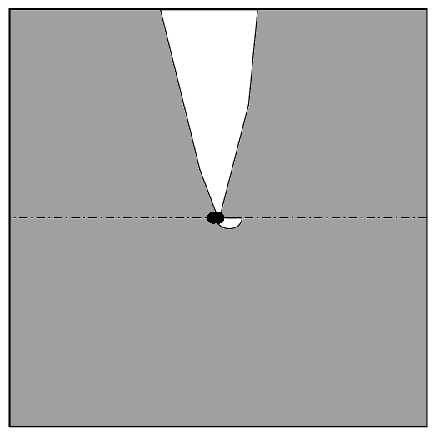}
    \includegraphics{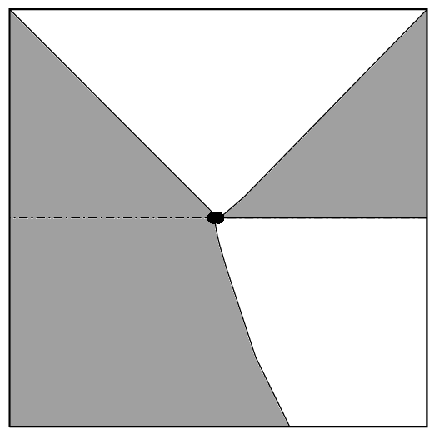}
    \includegraphics{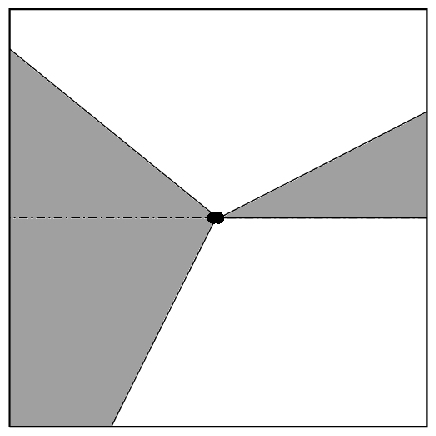}
  }
  \resizebox{\hsize}{!}{
    \raisebox{.23\hsize}{\parbox{1em}{B}}
    \includegraphics{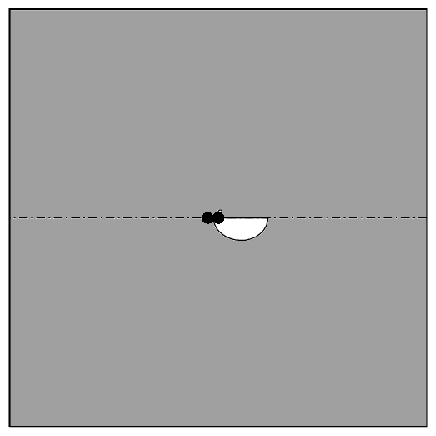}
    \includegraphics{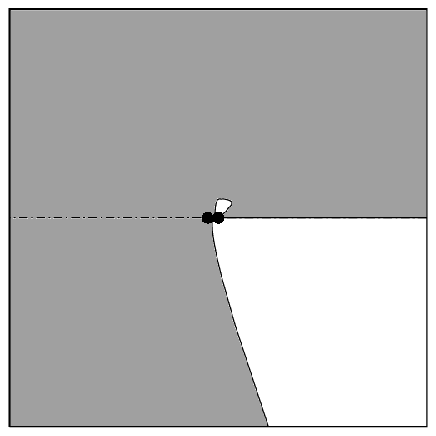}
    \includegraphics{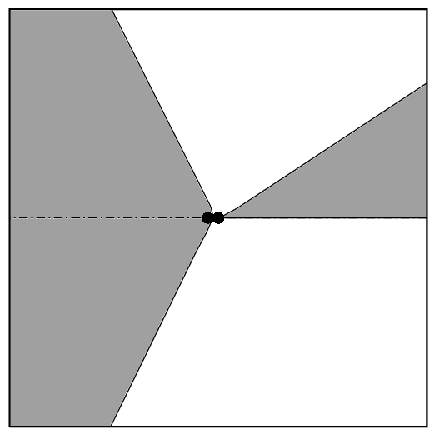}
  }
  \resizebox{\hsize}{!}{
    \raisebox{.23\hsize}{\parbox{1em}{C}}
    \includegraphics{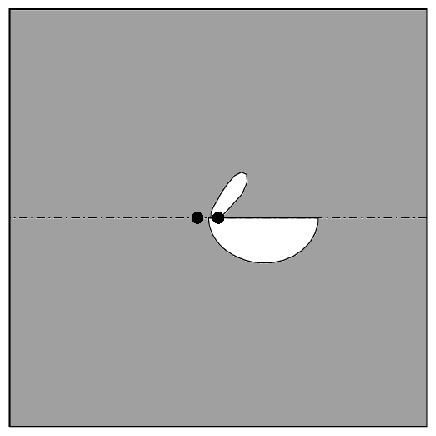}
    \includegraphics{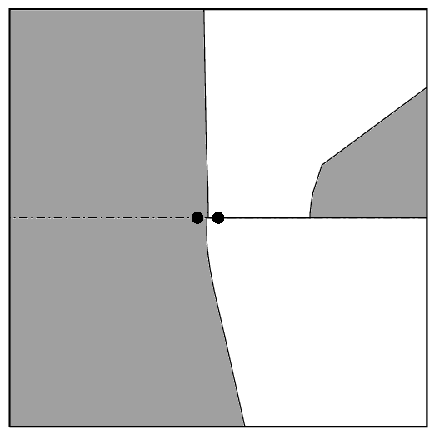}
    \includegraphics{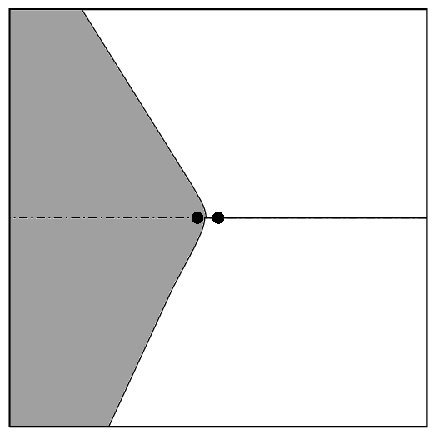}
  }
  \resizebox{\hsize}{!}{
    \hspace{1em}
    \parbox{0.3333\hsize}{\center \small $X=0.2$ }
    \parbox{0.3333\hsize}{\center \small $X=0.6$ }
    \parbox{0.3333\hsize}{\center \small $X=2.0$ }
  }
  \caption{A sample of \ion{H}{ii} region shapes. \ion{H}{ii} region boundary 
    is plotted in a plane which is perpendicular to the orbit and contains 
    the centers of both stars. In every frame our model (top) is compared to 
    its STB counterpart (bottom) obtained for the same value of $X$. The 
    columns illustrate the effect of increasing luminosity, (or, 
    equivalently, of increasing $X$). In each frame a region of 
    $80\times80\,\mathrm{AU}$ is shown; ionized regions are white.}
  \label{fig:hii_shapes}
\end{figure*}

The two sets of \ion{H}{ii} regions are compared in
Fig.~\ref{fig:hii_shapes}. The upper part of each frame shows our
model, and the lower one - the corresponding model obtained within the
framework of STB (note that all displayed models are symmetric with
respect to the orbital plane).

A clear general trend may be observed: for wide binaries and intense ionizing 
fluxes (see lower right corner of Fig.~\ref{fig:hii_shapes}) the shapes and 
sizes of our \ion{H}{ii} regions converge to those of STB. On the other hand, 
for close binaries and/or low ionizing luminosities, our \ion{H}{ii} regions 
become entirely different from those of STB: they are not axially symmetric, 
and their shapes may be fairly complicated (see Fig.~\ref{fig:hii_3d} and the 
left column of Fig.~\ref{fig:hii_shapes}). Another discrepancy concerns a 
large part of the space to the right of the secondary in 
Fig.~\ref{fig:hii_shapes}, which in our models is screened from UV photons by 
the dense disk around the secondary. This neutral region, best visible in the 
upper right frames of Fig.~\ref{fig:hii_shapes}, does not have its 
counterpart in STB models, and it only disappears when $L_\mathrm{ph}$ is 
large enough to ionize the disk. 

Note also that our \ion{H}{ii} regions are generally smaller than those of 
STB. This is due to the fact that the density of our RG wind is enhanced in 
the orbital plane compared to the spherically symmetric wind with the same 
mass loss rate and velocity. However, in the closest binary with the lowest 
$L_\mathrm{ph}$ our \ion{H}{ii} region becomes much larger than its STB 
counterpart. This is because the low-density funnels, which are very well 
developed in this case, are easily ionized, causing the \ion{H}{ii} region to 
evolve into a pair of semi-infinite jet-like lobes (upper left frame of 
Fig.~\ref{fig:hii_shapes}). 

\subsection{Spectra} 
\label{spectra}
The spherical grid centered on the secondary and containing the smoothed 
density distribution is also used to derive the {\it ff} spectra of our models.
Monochromatic volume emissivity of the ionized gas is calculated with the 
help of formulae taken from \cite{86Lang}. Subsequently, a viewing direction 
is chosen and the emissivity is integrated along the line of sight until the 
optical depth of 5 is reached (or across the whole \ion{H}{ii} region in 
optically thin cases). Observed monochromatic intensities are calculated for 
an assumed distance of $1\,\mathrm{kpc}$ between the observer and the 
radiation source.

\begin{figure*}
  \centering
  \resizebox{0.98\hsize}{!}{
    \includegraphics[angle=90]{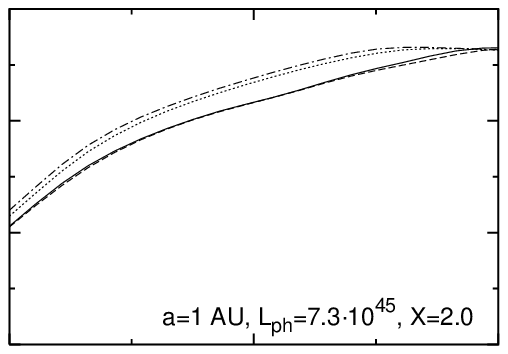}
    \includegraphics[angle=90]{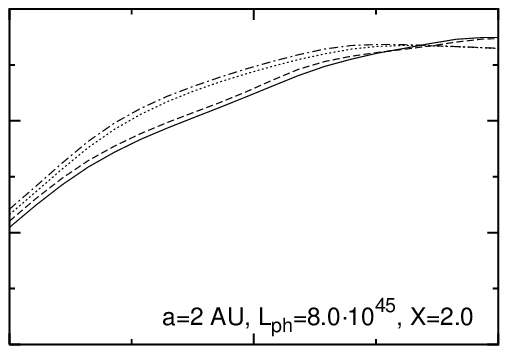}
    \includegraphics[angle=90]{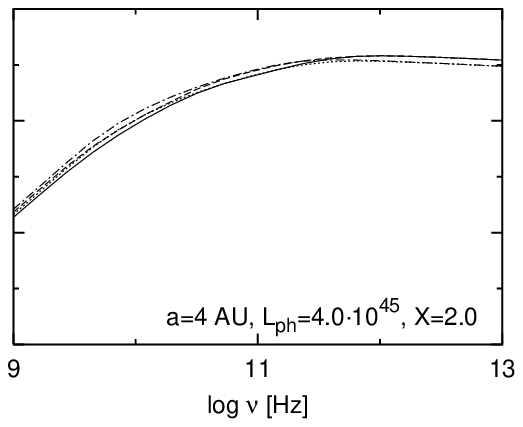}
  }
  \resizebox{0.98\hsize}{!}{
    \includegraphics[angle=90]{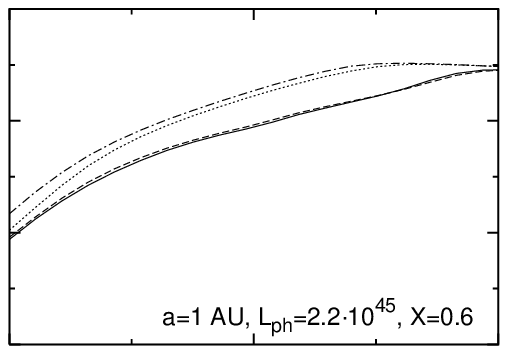}
    \includegraphics[angle=90]{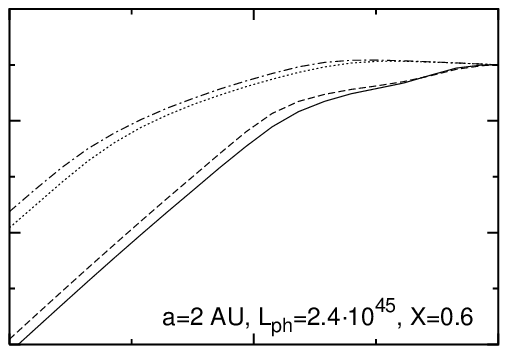}
    \includegraphics[angle=90]{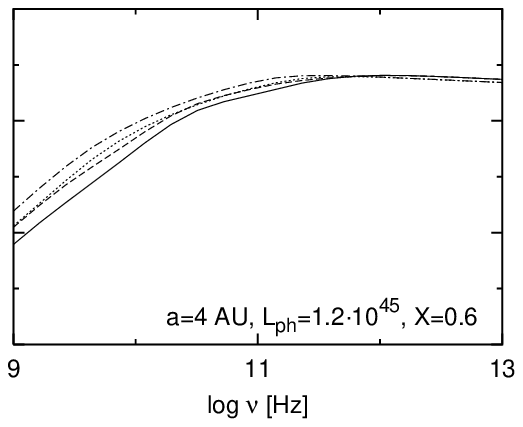}
  }
  \resizebox{0.98\hsize}{!}{
    \includegraphics[angle=90]{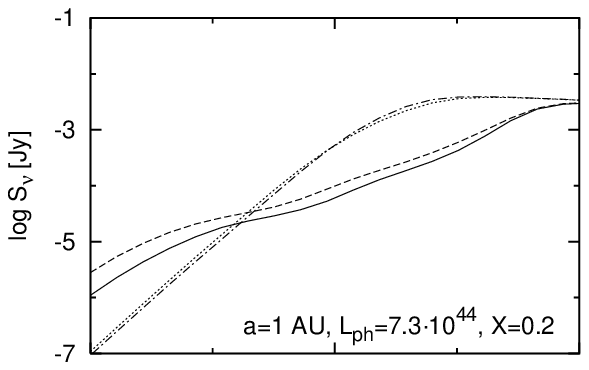}
    \includegraphics[angle=90]{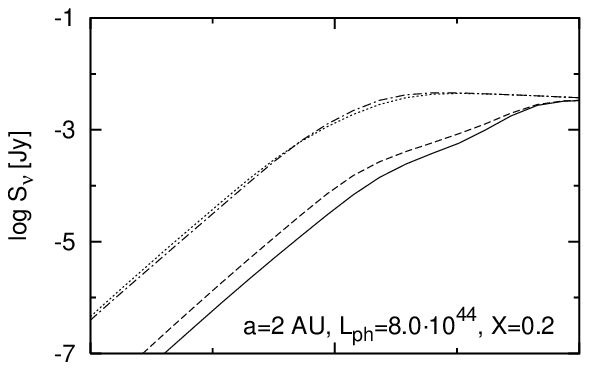}
    \includegraphics[angle=90]{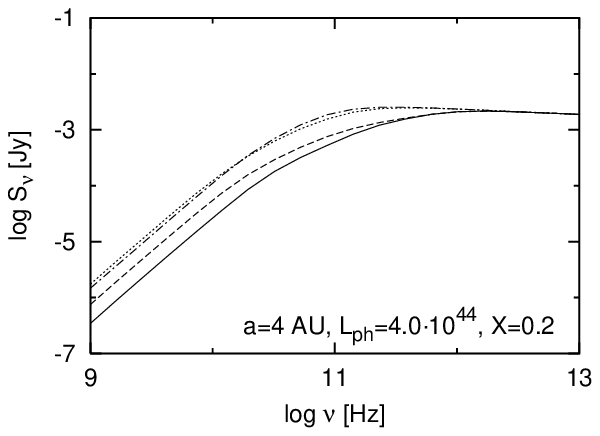}
  }
  \caption{Radio spectra of models shown in Fig.~\ref{fig:hii_shapes}. In all 
    cases $\dot{M}=10^{-7}$. Solid and dashed lines: our models viewed 
    pole-on and along the line joining the stars (with the primary in front 
    of the secondary), respectively. Dotted and dot-dashed lines: 
    corresponding STB models viewed pole-on and along the line joining the 
    stars, respectively.}
  \label{fig:hii_spectra}
\end{figure*}

\begin{figure} 
  \centering 
  \resizebox{\hsize}{!}{\includegraphics{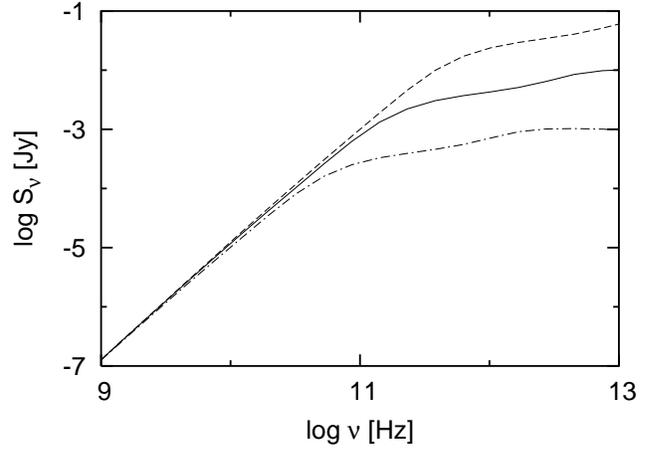}} 
  \caption{Scaling of the spectra. Compared are variants of model B with 
    $\dot{M}=10^{-7}\,M_{\sun}\,\mathrm{yr^{-1}},\ L_\mathrm{ph}=2.4\cdot10^{45}$ (solid line), 
    $\dot{M}=3\cdot10^{-7}\,M_{\sun}\,\mathrm{yr^{-1}},\ L_\mathrm{ph}=2.16\cdot10^{46}$ (dashed line) 
    and $\dot{M}=3\cdot10^{-8}\,M_{\sun}\,\mathrm{yr^{-1}},\ L_\mathrm{ph}=2.16\cdot10^{44}$ (dot-dashed line). 
    All three \ion{H}{ii} regions have the same shape and size. Their spectra 
    are practically identical, except that they are shifted by approximately 
    one order of magnitude in $S_\nu$ and a factor of $\sim 3$ in $\nu$ for 
    every factor of $\sim 3$ in $\dot{M}$ accompanied by one order of 
    magnitude increase in $L_\mathrm{ph}$.} 
  \label{fig:hii_spectrum_shift} 
\end{figure}

The resulting spectra of all models presented in Fig.~\ref{fig:hii_shapes} 
are displayed in Fig.~\ref{fig:hii_spectra} (note that in both figures the 
models are arranged in the same way, but the whole Fig.~\ref{fig:hii_spectra} 
is rotated counterclockwise by 90$\degr$). For comparison, the corresponding 
STB spectra are also shown. All spectra may easily be scaled to other values 
of $\dot M$ and $L_\mathrm{ph}$; the corresponding procedure is illustrated 
in Fig~\ref{fig:hii_spectrum_shift}.

As one might expect, in wider systems with high $L_\mathrm{ph}$ our spectra 
almost entirely converge to those of STB, while they are significantly 
different in close systems with low $L_\mathrm{ph}$. In such systems our 
spectral index may be smaller or larger than that of STB; the same is true 
for turnover frequency and intensity. To make things even more complicated, 
some spectra exhibit more than one turnover. This is because in some models
the distribution of density along the line of sight cannot be approximated by 
a single power law.

In general, the differences between the spectra of the same system viewed 
"pole-on" and along the line joining the stars (with the primary in front of 
the secondary) are larger in our models than in those of STB (slopes and 
turnover frequencies are however almost insensitive to such a change of 
viewing direction). On the other hand, our models predict weaker spectral 
variations with the orbital phase. This behaviour is explained by the fact 
that in our models the distinguished direction is perpendicular to the orbital 
plane, whereas the \ion{H}{ii} regions of STB are axially symmetric with 
respect to the line joining the stars.

\subsection{Radio maps} 
\label{maps}
\begin{figure*}
  \centering
  \resizebox{0.90\hsize}{!}{
    \includegraphics{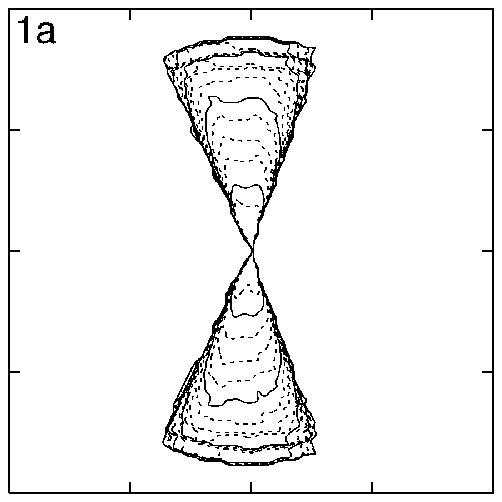}
    \includegraphics{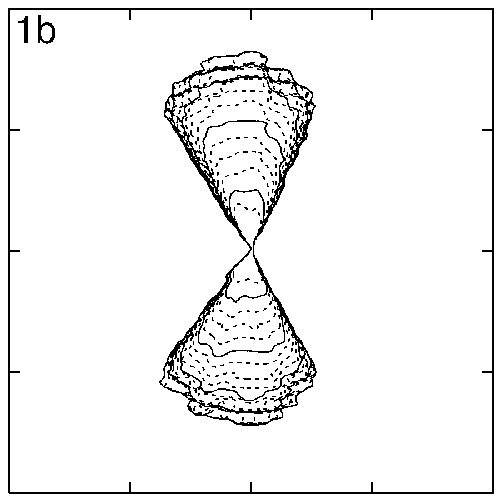}
    \includegraphics{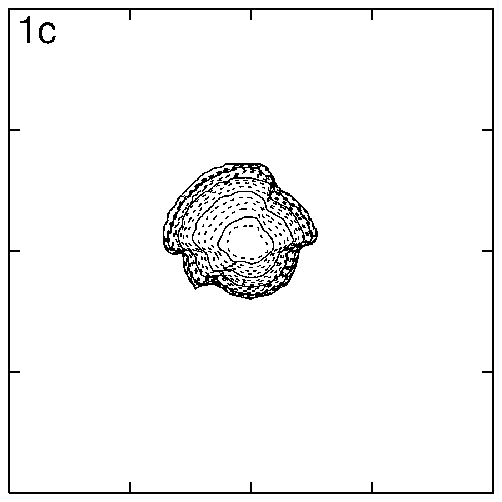}
  }
  \resizebox{0.90\hsize}{!}{
    \includegraphics{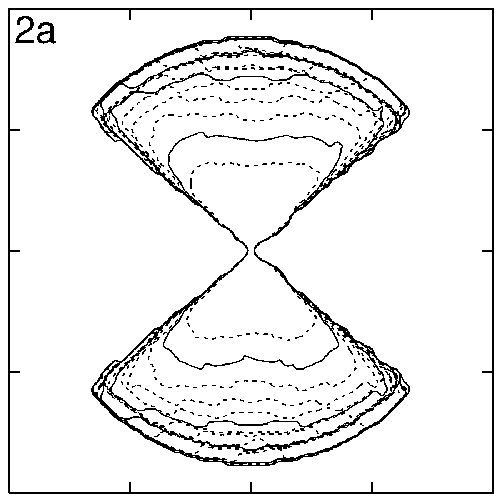}
    \includegraphics{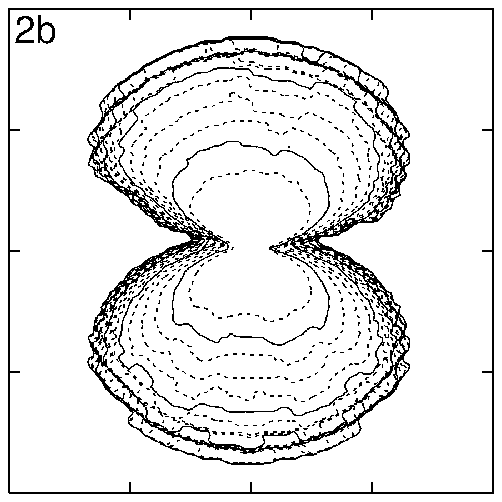}
    \includegraphics{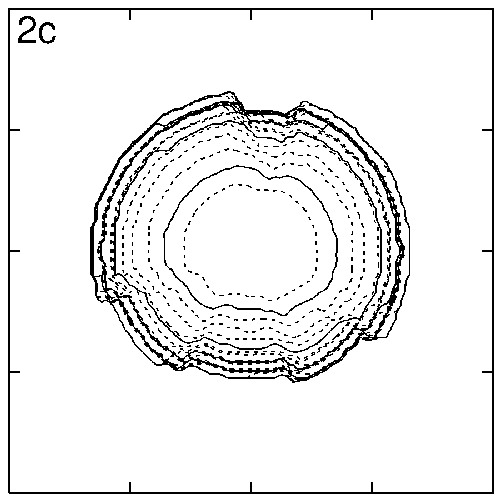}
  }
  \resizebox{0.90\hsize}{!}{
    \includegraphics{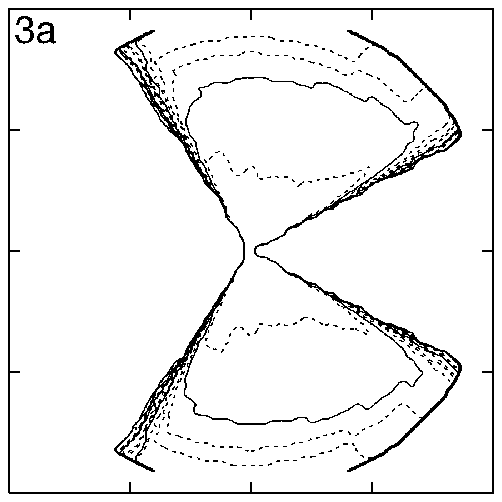}
    \includegraphics{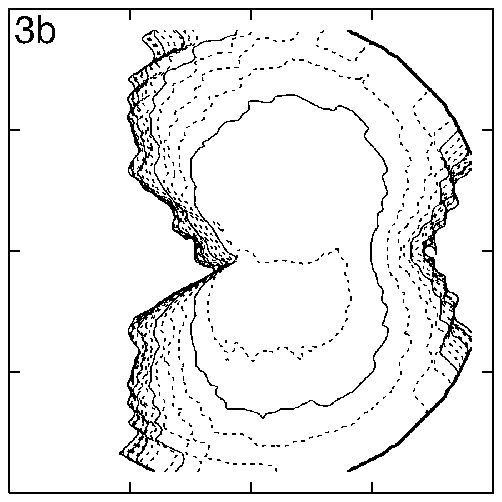}
    \includegraphics{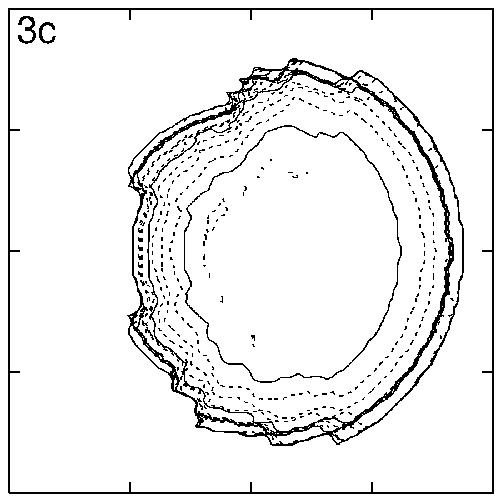}
  }
  \resizebox{0.90\hsize}{!}{
    \includegraphics{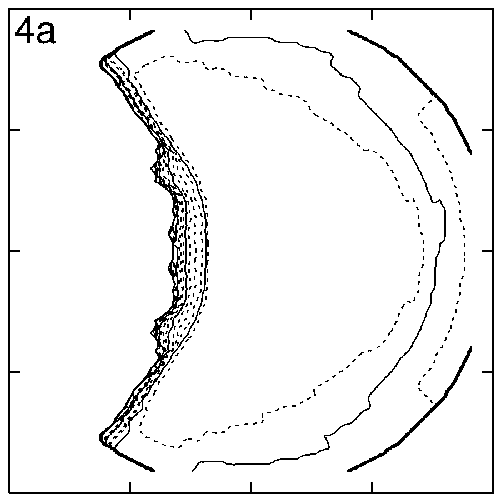}
    \includegraphics{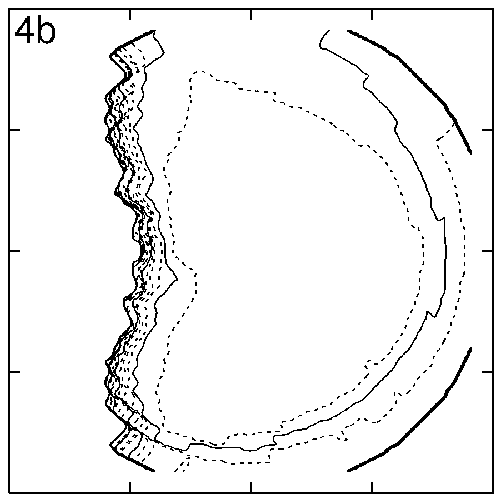}
    \includegraphics{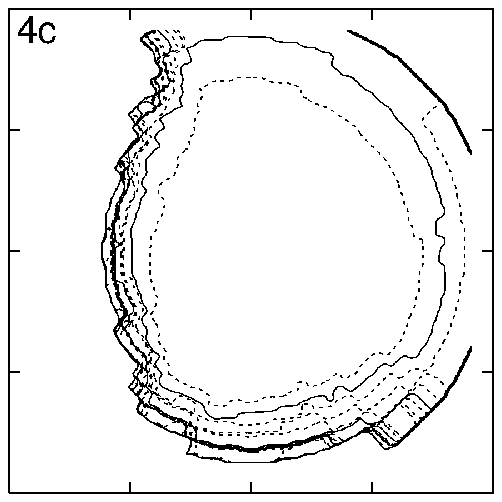}
  }
  \resizebox{0.90\hsize}{!}{
    \includegraphics{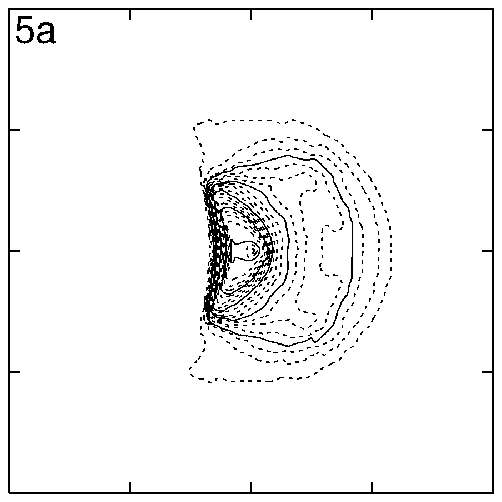}
    \includegraphics{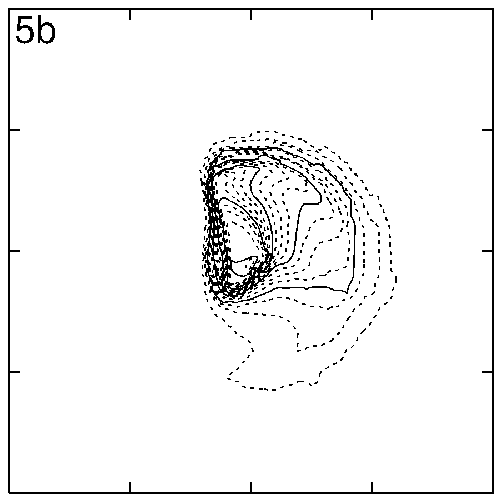}
    \includegraphics{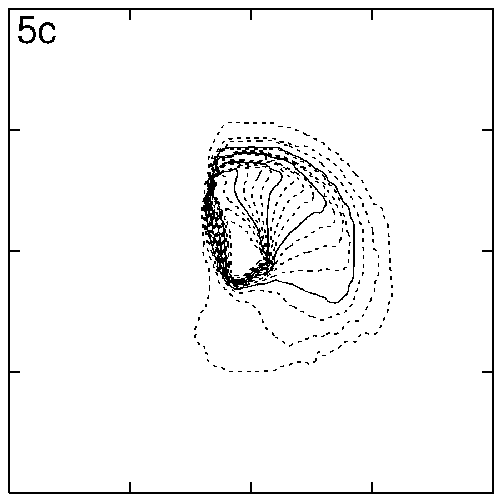}
  }
  \caption{Selected radiomaps at $5\,\mathrm{GHz}$. Radio intensity is shown 
    on a logarithmic scale, with solid contours plotted every order of 
    magnitude (except for frames {\bf 5a-c}, where a linear scale is used).
    Every frame covers an area of $80 \times 80\,\mathrm{AU}$. Left column: 
    side view (line of sight perpendicular to the line joining the stars in 
    the orbital plane). Right column: pole-on view (line of sight 
    perpendicular to the orbit). Middle column: intermediate case with line 
    of sight inclined $45\degr$ to the orbital plane. The jagged appearance 
    of some maps is an artifact due partly to limited resolution of the grid 
    on which the boundary of the \ion{H}{ii} region is defined. Model 
    parameters:
    {\bf 1a-c:} $a=1,\,\dot{M}=10^{-7},\,L_\mathrm{ph}=7.3\cdot10^{44},\,X=0.2$
    {\bf 2a-c:} $a=1,\,\dot{M}=10^{-7},\,L_\mathrm{ph}=2.2\cdot10^{45},\,X=0.6$
    {\bf 3a-c:} $a=2,\,\dot{M}=10^{-7},\,L_\mathrm{ph}=8.0\cdot10^{45},\,X=2.0$
    {\bf 4a-c:} $a=4,\,\dot{M}=10^{-7},\,L_\mathrm{ph}=4.0\cdot10^{45},\,X=2.0$
    {\bf 5a-c:} $a=4,\,\dot{M}=10^{-8},\,L_\mathrm{ph}=4.0\cdot10^{43},\,X=2.0$.
    Average radio luminosity integrated over frames is
    $0.01,
    \ 0.10,
    \ 0.20,
    \ 0.35
    \ \mathrm{and}\ 0.05\,\mathrm{mJy}$ for rows 1--5, respectively.
  }
  \label{fig:hii_maps}
\end{figure*}

Based on density distributions inside the ionized regions and the above
described integration technique we derive maps of monochromatic {\it ff} 
emission from our models at $5\,\mathrm{GHz}$. Depending on viewing angle and 
parameters of the binary, the maps display a rich variety of morphologies, of 
which a sample is shown in Fig.~\ref{fig:hii_maps}. In some frames the outer 
boundary of the RG wind model is visible as a regular circular contour. In 
such cases part of the emission is obviously not accounted for; however it 
never amounts to more that a few per cent of the total emission at 
$5\,\mathrm{GHz}$. Like in the case of spectra, the intensities indicated in
the maps correspond to a distance of $1\,\mathrm{kpc}$ between the observer 
and the symbiotic binary. The following observations can be made:

\begin{itemize}
  \item Bipolar structure with clearly separated lobes is a standard for 
    systems with $X\lesssim1$. 
  \item When the line of sight is inclined to the orbital plane, the lobes 
    may differ both in shape and intensity. The differences are primarily due 
    to the complicated shape of the effectively emitting ($\tau\sim1$) layer 
    in optically thick parts of the \ion{H}{ii} region (an example of such 
    region is shown in Fig.~\ref{fig:hii_3d}).
  \item A significant degree of asymmetry may also be seen in pole-on systems 
    (where only one lobe is visible). The primary cause of this asymmetry is 
    the 3-D spiral described in Sect.~\ref{shapes}.
\end{itemize}

The bipolar structure shown by our maps could be easily resolved by MERLIN 
after its upgrade is complete (e-MERLIN). Even now, MERLIN has sufficient 
resolution ($40\,\mathrm{mas}$ at $5\,\mathrm{GHz}$ where it is most 
sensitive) but its sensitivity is too low. However, e-MERLIN, with its new 15 
and $22\,\mathrm{GHz}$ receivers will be able to perform imaging at a 
resolution of $\sim 8\,\mathrm{AU}$ at $1\,\mathrm{kpc}$ for sources as weak 
as a few of $\mu\mathrm{Jy}$.

\section{Discussion}
\label{discussion}

We have demonstrated that the STB model is not applicable to systems with 
$a\la 4$ and $X\la 2$. In such systems the \ion{H}{ii} region is not axially 
symmetric, and in most cases it is much smaller than its STB counterpart. 
This is because the basic STB assumption of a spherically symmetric RG wind 
is violated. Even if the wind is emitted uniformly from the whole surface of 
the red giant, it immediately loses its original symmetry due to 
gravitational attraction of the secondary. As a result, a large-scale density 
enhancement develops in the orbital plane, with many smaller scale 
substructures (funnels, spirals and a compact circumsecondary disk). Factors 
not accounted for in our simulations (RG wind originally enhanced in the 
orbital plane and/or outflow through the inner Lagrangian point) can only 
amplify this effect. Based on Fig.~\ref{fig:hii_shapes} the following 
qualitative predictions can be made:
\begin{itemize}
  \item In most S-type systems large amounts of neutral material should be 
    observed at the orbital plane. This is because the circumsecondary disk 
    and the orbital density enhancement are much harder to ionize than the 
    spherically symmetric RG wind. This prediction holds for the range of 
    parameters investigated here 
    ($\dot{M}\sim10^{-7}\,\mathrm{M}_{\sun}\,\mathrm{yr}^{-1}$; $L_\mathrm{ph}\la 10^{46}\,\mathrm{phot\,s^{-1}}$) 
    as well as for $\dot{M}$ and $L_\mathrm{ph}$ scaled as explained in 
    Fig.~\ref{fig:hii_spectrum_shift}.
  \item Both the brightness and spectra of the radio continuum emission from 
    edge-on systems should show little variability with the orbital phase. 
    This is because the preferred direction (in which the ionizing quanta 
    propagate most easily) is perpendicular to the orbital plane. Note that 
    in the STB model the preferred direction is the line joining the stars 
    (or rather its extension beyond the secondary).
  \item High resolution imaging of edge-on systems with lower $L_\mathrm{ph}$
    should reveal two maxima of radio emission. In some systems traces of 
    spiral structure may be detected.
\end{itemize}

In fact, the first prediction has already been observationally
verified. For many S-type systems there are strong indications for low
values of $X$ \citep[$\la 1/3$;][]{91MNSV} and/or \ion{H}{ii} which
are ionization-bounded at least in the orbital plane (\cite{ibm94}. In
accordance with Fig.~\ref{fig:hii_shapes}, such nebulae have also been
observed for larger $X$'s. Recently, \cite{02Q} have shown that in the 
case of \object{AR Pav}, the central absorption in $\mathrm{H}_\alpha$ 
is formed in a neutral portion of the RG wind which is strongly 
concentrated towards the orbital plane, (with the total hydrogen column 
density $N(\mathrm{H})\ga \mathrm{a\ few} \times 10^{21}\,\mathrm{cm}^{-2}$).

In systems with $a\la2$ and $X\la 1$ our radio spectra significantly
diverge from those of STB. First, the location and extent of the
transition region between optically thin and optically thick parts of
the spectrum are different. Second, the shape of the transition region
is more complicated in our models. Third, our models radiate much less
energy in the optically thick part of the spectrum than their STB
counterparts (the difference can amount to two orders of magnitude).

The latter result can be understood as a simple consequence of the
smaller spatial extent of our \ion{H}{ii} regions (the optically thick
flux is roughly proportional to the projected surface area of the
ionized part of the wind). In principle, the remaining predictions
could be verified by spectroscopic observations. Unfortunately,
observational data, especially at higher frequencies, are still too
poor to discriminate between various theoretical models. The broadest
survey available, reported by \cite{mio02,mio02lp} is based on
mm/submm observations of a sample of 20 S-type systems in
quiescence. Radio emission from these systems is found to be optically
thick at least up to $\sim 1.3\,\mathrm{mm}\ (243\,\mathrm{GHz} =
\nu_t$ = turnover frequency). This is compatible both with the STB
model, which for typical quiescent S-type systems predicts
$\nu_\mathrm{t}\ga 300\,\mathrm{GHz}$ \citep{93SKT}, and with our
simulations (see Fig.~\ref{fig:hii_spectra}).

At present the only system which allows for a quantitative analysis of
the spectrum is
\object{CI Cyg}, where $\nu_\mathrm{t}\approx 30\,\mathrm{GHz}$. For
the STB model the outcome of such an analysis is rather
unfavourable. In their framework the orbital separation may be
estimated from
\begin{equation}
a=300\bigg(\frac{T_\mathrm{e}}{10^4\,\mathrm{K}}\bigg)^{-\frac{1}{2}}
     \bigg(\frac{\nu_\mathrm{t}}{\mathrm{GHz}}\bigg)^{-1}
     \bigg(\frac{S_\mathrm{t}}{\mathrm{mJy}}\bigg)^{\frac{1}{2}}
     \bigg(\frac{d}{\mathrm{kpc}}\bigg){\,\mathrm{AU}},
     \label{eq:stb_binsep}
\end{equation}
where $T_\mathrm{e}$ is the electron temperature of the ionized wind,
and $S_\mathrm{t}$ is the flux received at the turnover
frequency. This estimate should be accurate within a factor of 2 for a
range of $X$ covering two orders of magnitude. However, in the case of
\object{CI Cyg} it overestimates $a$ by a factor of 35
($70\,\mathrm{AU}$ compared to $\sim2\,\mathrm{AU}$ obtained from the
spectroscopic orbit solution. On the other hand, if one inserts $a=2$
in (\ref{eq:stb_binsep}) and calculates the turnover flux, then the
derived $S_\mathrm{t}$ is $10^3$ times lower than the actually
observed one.

The STB framework also provides an estimate for $L_\mathrm{ph}$
\begin{eqnarray}
   L_\mathrm{ph}&=&10^{56}\alpha_\mathrm{rec}
   \bigg(\frac{T_\mathrm{e}}{10^4\,\mathrm{K}}\bigg)^{-0.35}
   \bigg(\frac{\nu}{\mathrm{GHz}}\bigg)^{0.1}\nonumber\\
   &&\bigg(\frac{S_\mathrm{\nu}}{\mathrm{mJy}}\bigg)
   \bigg(\frac{d}{\mathrm{kpc}}\bigg)^2{\,\mathrm{phot}\,\mathrm{s}^{-1}},
   \label{eq:stb_ion_lum}
\end{eqnarray}
where $S_\mathrm{\nu}$ is the flux received in the optically thin part of the 
spectrum at frequency $\nu$ \cite{95is}. However, for \object{CI Cyg} formula 
(\ref{eq:stb_ion_lum}) yields a value 20 times lower than that obtained from 
optical/UV \ion{H}{i} and \ion{He}{ii} emission lines and  {\it bf+ff} 
continuum. 

Other inconsistencies between the observed spectrum of \object{CI Cyg} and 
the STB model are discussed by \cite{01MI}, who suggest that the STB approach 
might not apply to this system because it is likely that the red giant fills 
or nearly fills its Roche lobe, and the mass loss is concentrated in a stream 
flowing through the inner Lagrangian point. If this is really the case, then 
our models, which predict concentration of the outflow in the equatorial 
plane and formation of the circumsecondary disk, should provide a better 
approximation of the observed spectrum than that obtained within the STB 
framework. 

{  Unfortunately, in this particular case they do not perform much better. 
A glance at Fig.~\ref{fig:hii_spectra} indicates that in order to obtain 
$\nu_\mathrm{t}$ as low as $\sim30\,\mathrm{GHz}$ we would have to assume an 
unacceptably low $L_\mathrm{ph}$ and/or $\dot M$ much lower than 
$10^{-7}\,M_{\sun}\,\mathrm{yr}^{-1}$. } A likely solution to this problem is 
suggested by the spectrum for $a=1\,\mathrm{AU},\ X=0.2$ in 
Fig.~\ref{fig:hii_spectra}, which is the only one showing an excess in the 
optically thick part. With increasing $\dot M$ and $L_\mathrm{ph}$ the 
excess would shift to higher frequencies, mimicking the turnover observed in 
\object{CI Cyg}. The fact that the spectrum obtained for $a=1\,\mathrm{AU}$ 
offers a better fitting possibility for the system with $a=2\,\mathrm{AU}$ 
seems to indicate that our model of the latter underestimates the density of 
the circumsecondary disk and/or overestimates the density of the funnels. 
Such a conclusion is compatible with arguments concerning the outflow through 
the inner Lagrangian point. 
 
The deciding tests of our results  will become possible when high-resolution 
radio maps of S-type systems are obtained. Our results suggest that S-type 
systems will be interesting targets for e-MERLIN and ALMA.

While the models presented here are closer to reality than those obtained 
within the STB approach, there is still a considerable room for improvements. 
We have only marginally resolved the disk around the secondary, whose shape 
and extent may strongly influence shape and extent of the \ion{H}{ii} region. 
To improve on that, future work should include effects of radiation transfer 
and viscosity, increase the mass--resolution of the SPH code and, 
accordingly, decrease its time--step. Note, however, that with radiation 
transfer and viscosity included, the simple scaling relation presented in 
Fig.~\ref{fig:hii_spectrum_shift} would not work. This means that RG wind 
models extending to hundreds of $\mathrm{AU}$ would have to be obtained in 
order to generate \ion{H}{ii} regions emitting as much energy in the 
optically thin regime as the observed ones. Such simulations would push the 
CPU requirements beyond currently acceptable limits. When they are hopefully 
done in the future and a reliable thermal structure of the flow is obtained, 
the thermal ionization should be taken into account, which at least in some 
cases may significantly contribute to the radio spectra. 
 
\begin{acknowledgements}
This research was supported by the Committee for Scientific Research through 
grants 2P03D~014~19 and 5P03D~019~20. 
\end{acknowledgements}

\bibliographystyle{aa}
\bibliography{symbiotic_hii.bib}

\end{document}